\documentclass[conference]{IEEEtran}

\usepackage{graphicx}
\usepackage{amsmath}
\begin{document}
%
% paper title
% can use linebreaks \\ within to get better formatting as desired
\title{Intercept Probability Analysis of Cooperative Wireless Networks with Best Relay Selection in the Presence of Eavesdropping Attack}
%
%
% author names and IEEE memberships
% note positions of commas and nonbreaking spaces ( ~ ) LaTeX will not break
% a structure at a ~ so this keeps an author's name from being broken across
% two lines.
% use \thanks{} to gain access to the first footnote area
% a separate \thanks must be used for each paragraph as LaTeX2e's \thanks
% was not built to handle multiple paragraphs
%
\author{\IEEEauthorblockN{Yulong Zou$^{*,\dag}$, Xianbin Wang$^{*}$, and Weiming Shen$^{*,\dag}$}
\IEEEauthorblockA{
$^{*}$Electrical \& Computer Engineering Department, The University of Western Ontario, London, Ontario, Canada\\
$^{\dag}$College of Electronics and Information Engineering, Tongji University, Shanghai, China\\
Email: \{yzou45, xianbin.wang, wshen\}@uwo.ca
}
}

% make the title area
\maketitle

\begin{abstract}
Due to the broadcast nature of wireless medium, wireless communication is extremely vulnerable to eavesdropping attack. Physical-layer security is emerging as a new paradigm to prevent the eavesdropper from interception by exploiting the physical characteristics of wireless channels, which has recently attracted a lot of research attentions. In this paper, we consider the physical-layer security in cooperative wireless networks with multiple decode-and-forward (DF) relays and investigate the best relay selection in the presence of eavesdropping attack. For the comparison purpose, we also examine the conventional direct transmission without relay and traditional max-min relay selection. We derive closed-form intercept probability expressions of the direct transmission, traditional max-min relay selection, and proposed best relay selection schemes in Rayleigh fading channels. Numerical results show that the proposed best relay selection scheme strictly outperforms the traditional direct transmission and max-min relay selection schemes in terms of intercept probability. In addition, as the number of relays increases, the intercept probabilities of both traditional max-min relay selection and proposed best relay selection schemes decrease significantly, showing the advantage of exploiting multiple relays against eavesdropping attack.

\end{abstract}


\IEEEpeerreviewmaketitle

\section{Introduction}
% The very first letter is a 2 line initial drop letter followed
% by the rest of the first word in caps.
%
% form to use if the first word consists of a single letter:
% \IEEEPARstart{A}{demo} file is ....
%
% form to use if you need the single drop letter followed by
% normal text (unknown if ever used by IEEE):
% \IEEEPARstart{A}{}demo file is ....
%
% Some journals put the first two words in caps:
% \IEEEPARstart{T}{his demo} file is ....
%
% Here we have the typical use of a "T" for an initial drop letter
% and "HIS" in caps to complete the first word.
\IEEEPARstart In wireless networks, radio signals can be overheard by unauthorized users due to the broadcast nature of wireless medium, which makes the wireless communication systems vulnerable to eavesdropping attack. Secret key encryption techniques have been widely used to prevent eavesdropping and ensure the confidentiality of signal transmissions. However, the cryptographic techniques rely on secret keys and introduce additional complexities due to the dynamic distribution and management of secret keys. To this end, physical-layer security is emerging as an alternative paradigm to prevent the eavesdropper attack and assure the secure communication by exploiting the physical characteristics of wireless channels. The physical-layer security work was pioneered by Wyner [1] and further extended in [2], where an information-theoretic framework has been established by developing achievable secrecy rates. It has been proven in [2] that in the presence of an eavesdropper, a so-called \emph{secrecy capacity} is shown as the difference between the channel capacity from source to destination (called main link) and that from source to eavesdropper (called wiretap link). If the secrecy capacity is negative, the eavesdropper can intercept the transmission from source to destination and an intercept event occurs in this case. Due to the wireless fading effect, the secrecy capacity is severely limited, which results in an increase in the intercept probability. To alleviate this problem, some existing work is proposed to improve the secrecy capacity by taking advantage of multiple antennas [3] and [4].

However, it may be difficult to implement multiple antennas in some cases (e.g., handheld terminals, sensor nodes, etc.) due to the limitation in physical size and power consumption. As an alternative, user cooperation is proposed as an effective means to combat wireless fading, which also has great potential to improve the secrecy capacity of wireless transmissions in the presence of eavesdropping attack. In [5], the authors studied the secrecy capacity of wireless transmissions in the presence of an eavesdropper with a relay node, where the amplify-and-forward (AF), decode-and-forward (DF), and compress-and-forward (CF) relaying protocols are examined and compared with each other. The cooperative jamming was proposed in [6] by allowing multiple users to cooperate with each other in preventing eavesdropping and analyzed in terms of the achievable secrecy rate. In [7], the cooperation strategy was further examined to enhance the physical-layer security and a so-called noise-forwarding scheme was proposed, where the relay node attempts to send codewords independent of the source message to confuse the eavesdropper. In addition, in [8] and [9], the authors explored the cooperative relays for physical-layer security improvement and developed the corresponding secrecy capacity performance, showing that the cooperative relays can significantly increase the secrecy capacity.

In this paper, we consider a cooperative wireless network with multiple DF relays in the presence of an eavesdropper and examine the best relay selection to improve wireless security against eavesdropping attack. Differing from the traditional max-min relay selection criterion in [10] where only the channel state information (CSI) of two-hop relay links (i.e., source-relay and relay-destination) are considered, we here have to take into account additional CSI of the eavesdropper's links, in addition to the two-hop relay links' CSI. The main contributions of this paper are summarized as follows. First, we propose the best relay selection scheme in a cooperative wireless networks with multiple DF relays in the presence of eavesdropping attack. We also examine the direct transmission without relay and traditional max-min relay selection as benchmark schemes. Secondly, we derive closed-form expressions of intercept probability for the direct transmission, traditional max-min relay selection, and proposed best relay selection schemes in Rayleigh fading channels.

The remainder of this paper is organized as follows. Section II presents the system model and describes the direct transmission, traditional max-min relay selection, and proposed best relay selection schemes. In Section III, we derive closed-form intercept probability expressions of the direct transmission, traditional max-min relay selection, and proposed best relay selection schemes over Rayleigh fading channels. In Section IV, we conduct numerical intercept probability evaluation to show the advantage of proposed best relay selection over traditional max-min relay selection. Finally, we make some concluding remarks in Section V.

\section{System Model and Proposed Best Relay Selection Scheme}

\subsection{System Model}
\begin{figure}
  \centering
  {\includegraphics[scale=0.65]{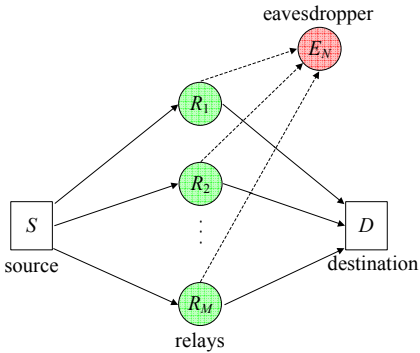}\\
  \caption{A cooperative wireless network consisting of multiple decode-and-forward (DF) relays in the presence of an eavesdropper.}\label{Fig1}}
\end{figure}

Consider a cooperative wireless network consisting of one source, one destination, and $M$ DF relays in the presence of an eavesdropper as shown in Fig. 1, where all nodes are equipped with single antenna and the solid and dash lines represent the main and wiretap links, respectively. The main and wiretap links both are modeled as Rayleigh fading channels and the thermal noise received at any node is modeled as a complex Gaussian random variable with zero mean and variance $\sigma^2_n$, i.e., ${\mathcal{CN}}(0,\sigma^2_n)$. Following [8], we consider that $M$ relays are exploited to assist the transmission from source to destination and the direct links from source to destination and eavesdropper are not available, e.g., the destination and eavesdropper both are out of the coverage area. For notational convenience, $M$ relays are denoted by ${\cal{R}} = \{ {{R}}_i |i = 1,2, \cdots ,M\}$. Differing from the existing work [8] in which all relays participate in forwarding the source messages to destination, we here consider the use of the best relay only to forward the message transmission from source to destination. More specifically, the source node first broadcasts the message to cooperative relays among which only the best relay will be selected to forward its received signal to destination. Meanwhile, the eavesdropper monitors the transmission from the best relay to destination and attempts to interpret the source message. Following [8], we assume that the eavesdropper knows everything about the signal transmission from source via relay to destination, including the encoding scheme at source, forwarding protocol at relays, and decoding method at destination, except that the source signal is confidential.

It needs to be pointed out that in order to effectively prevent the eavesdropper from interception, not only the CSI of main links, but also the wiretap links' CSI should be considered in the best relay selection. This differs from the traditional relay selection in [10] where only the two-hop relay links' CSI is considered in performing relay selection. Similarly to [8], we here assume that the global CSI of both main and wiretap links are available, which is a common assumption in the physical-layer security literature. Notice that the wiretaps link's CSI can be estimated and obtained by monitoring the eavesdropper's transmissions as discussed in [8]. In the following, we first describe the conventional direct transmission without relay and then present the traditional max-min relay selection and proposed best relay selection schemes.

\subsection{Direct Transmission}
For comparison purpose, this subsection describes the conventional direct transmission without relay. Consider that the source transmits a signal $s$ ($E(|s{|^2}) = 1$) with power $P$. Thus, the received signal at destination is expressed as
\begin{equation}\label{equa1}
r_d=\sqrt{P}h_{sd}s+n_d,
\end{equation}
where $h_{sd}$ represents a fading coefficient of the channel from source to destination and $n_d \sim {\mathcal{CN}}(0,\sigma^2_n)$ represents additive white Gaussian noise (AWGN) at destination. Meanwhile, due to the broadcast nature of wireless transmission, the eavesdropper also receives a copy of the source signal $s$ and the corresponding received signal is written as
\begin{equation}\label{equa2}
r_e=\sqrt{P}h_{se}s+n_e,
\end{equation}
where $h_{se}$ represents a fading coefficient of the channel from source to eavesdropper and $n_e \sim {\mathcal{CN}}(0,\sigma^2_n)$ represents AWGN at eavesdropper. Assuming the optimal Gaussian codebook used at source, the maximal achievable rate (also known as channel capacity) of the direct transmission from source to destination is obtained from Eq. (1) as
\begin{equation}\label{equa3}
C^{\textrm{direct}}_{sd}=\log_2(1+\dfrac{|h_{sd}|^2P}{\sigma^2_n}),
\end{equation}
where ${\sigma^2_n}$ is the noise variance. Similarly, from Eq. (2), the capacity of wiretap link from source to eavesdropper is easily given by
\begin{equation}\label{equa4}
C^{\textrm{direct}}_{se}=\log_2(1+\dfrac{|h_{se}|^2P}{\sigma^2_n}).
\end{equation}
It has been proven in [2] that the secrecy capacity is shown as the difference between the capacity of main link and that of wiretap link. Hence, the secrecy capacity of direct transmission is given by
\begin{equation}\label{equa5}
C^{\textrm{direct}}_{s}=C^{\textrm{direct}}_{sd} - C^{\textrm{direct}}_{se},
\end{equation}
where $C^{\textrm{direct}}_{sd}$ and $C^{\textrm{direct}}_{se}$ are given in Eqs. (3) and (4), respectively. As discussed in [2], when the secrecy capacity is negative (i.e., the capacity of main link falls below the wiretap link's capacity), the eavesdropper can intercept the source signal and an intercept event occurs. Thus, the probability that the eavesdropper successfully intercepts source signal, called \emph{intercept probability}, is a key metric in evaluating the performance of physical-layer security. In this paper, we mainly focus on how to improve the intercept probability by exploiting the best relay selection to enhance the wireless security against eavesdropping. The following subsection describes the relay selection for physical-layer security in the presence of eavesdropper attack.

\subsection{Relay Selection}
Considering the DF protocol, the relay first decodes its received signal from source and then re-encodes and transmits its decoded outcome to the destination. More specifically, the source node first broadcasts the signal $s$ to $M$ relays that attempt to decode their received signals. Then, only the best relay is selected to re-encode and transmit its decoded outcome to the destination. Notice that in the DF relaying transmission, the source signal $s$ is transmitted twice from the source and relay, respectively. In order to make a fair comparison with the direct transmission, the total amount of transmit power at source and relay shall be limited to $P$. By using the equal-power allocation for simplicity, the transmit power at source and relay is given by $P/2$. In DF relaying transmission, either source-relay or relay-destination links in failure will result in the two-hop DF transmission in failure, implying that the capacity of DF transmission is the minimum of the capacity from source to relay and that from relay to destination. Hence, considering $R_i$ as the best relay, we can obtain the capacity of DF relaying transmission from source via $R_i$ to destination as
\begin{equation}\label{equa6}
C^{\textrm{DF}}_{sid}=\min({C_{si},C_{id}}),
\end{equation}
where $C_{si}$ and $C_{id}$, respectively, represent the channel capacity from source to $R_i$ and that from $R_i$ to destination, which are given by
\begin{equation}\label{equa7}
C_{si}=\log_2(1+\dfrac{|h_{si}|^2P}{2\sigma^2_n}),
\end{equation}
and
\begin{equation}\label{equa8}
C_{id}=\log_2(1+\dfrac{|h_{id}|^2P}{2\sigma^2_n}).
\end{equation}
Meanwhile, the eavesdropper can overhear the transmission from $R_i$ to destination. Hence, the channel capacity from $R_i$ to eavesdropper can be easily obtained as
\begin{equation}\label{equa9}
C^{\textrm{DF}}_{ie}=\log_2(1+\dfrac{|h_{ie}|^2P}{2\sigma^2_n}).
\end{equation}
Combining Eqs. (6) and (9), the secrecy capacity of DF relaying transmission with $R_i$ is given by
\begin{equation}\label{equa10}
\begin{split}
C^{\textrm{DF}}_{i}=&C^{\textrm{DF}}_{sid} - C^{\textrm{DF}}_{ie}\\
=&{\log _2}\left(1 + \frac{{\min ( |{h_{si}}{|^2},|{h_{id}}{|^2}) P}}{{2\sigma _n^2}}\right) \\
&- {\log _2}\left(1 + \frac{{|{h_{ie}}{|^2}P}}{{2\sigma _n^2}}\right).
\end{split}
\end{equation}
The following presents the traditional max-min relay selection and proposed best relay selection schemes.

\subsubsection{Traditional Max-Min Relay Selection Scheme}Let us first present the traditional max-min relay selection scheme for the comparison purpose. In the traditional relay selection scheme, the relay that maximizes the capacity of DF relaying transmission $C^{\textrm{DF}}_{sid}$ is viewed as the best relay. Thus, the traditional relay selection criterion is obtained from Eq. (6) as
\begin{equation}\label{equa11}
\begin{split}
\textrm{BestRelay} &= \arg \mathop {\max }\limits_{i \in {\cal{R}}}C^{\textrm{DF}}_{sid}\\
&= \arg \mathop {\max }\limits_{i \in {\cal{R}}}\min({|h_{si}|^2,|h_{id}|^2}),
\end{split}
\end{equation}
which is the traditional max-min relay selection criterion as given by Eq. (1) in [10]. As shown in Eq. (11), only the main links' CSI $|h_{si}|^2$ and $|h_{id}|^2$ is considered in the max-min relay selection scheme without considering the eavesdropper's CSI $|h_{ie}|^2$.

\subsubsection{Proposed Best Relay Selection Scheme}We now propose the best relay selection criterion considering the CSI of both main and wiretap links, in which the relay that maximizes the secrecy capacity of DF relaying transmission is selected as the best relay. Thus, the best relay selection criterion is obtained from Eq. (10) as
\begin{equation}\label{equa12}
\begin{split}
\textrm{BestRelay} &= \arg \mathop {\max }\limits_{i \in {\cal{R}}}C^{\textrm{DF}}_{i}\\
&=\arg \mathop {\max }\limits_{i \in {\cal{R}}} \frac{{\min ( |{h_{si}}{|^2},|{h_{id}}{|^2}) P + 2\sigma _n^2}}{{|{h_{ie}}{|^2}P + 2\sigma _n^2}}.
\end{split}
\end{equation}
One can observe from Eq. (12) that the proposed best relay selection scheme takes into account not only the main links' CSI $|h_{si}|^2$ and $|h_{id}|^2$, but also the wiretap link's CSI $|h_{ie}|^2$. This differs from the traditional max-min relay selection criterion in Eq. (11) where only the main links' CSI is considered. Notice that the transmit power $P$ in Eq. (12) is a known parameter and the noise variance $\sigma _n^2$ is shown as $\sigma _n^2 =\kappa TB$ [12], where $\kappa $ is Boltzmann constant (i.e., $\kappa  = 1.38 \times {10^{ - 23}}$), $T$ is room temperature, and $B$ is system bandwidth. Since the room temperature $T$ and system bandwidth $B$ both are predetermined, the noise variance $\sigma_n^2$ can be easily obtained. It is pointed out that using the proposed best relay selection criterion in Eq. (12), we can further develop a centralized or distributed relay selection algorithm. To be specific, for a centralized relay selection, the source node needs to maintain a table that consists of $M$ relays and related CSI (i.e., $|h_{si}|^2$, $|h_{id}|^2$ and $|h_{ie}|^2$). In this way, the best relay can be easily determined by looking up the table using the proposed criterion in Eq. (12), which is referred to as centralized relay selection strategy. For a distributed relay selection, each relay maintains a timer and sets an initial value of the timer in inverse proportional to $\frac{{\min ( |{h_{si}}{|^2},|{h_{id}}{|^2}) P + 2\sigma _n^2}}{{|{h_{ie}}{|^2}P + 2\sigma _n^2}}$ in Eq. (12), resulting in the best relay with the smallest initial value for its timer. As a consequence, the best relay exhausts its timer earliest compared with the other relays, and then broadcasts a control packet to notify the source node and other relays [11].

\section{Intercept Probability Analysis}
In this section, we derive closed-form intercept probability expressions of the direct transmission, traditional max-min relay selection and proposed best relay selection schemes over Rayleigh fading channels.

\subsection{Direct Transmission}
Let us first analyze the intercept probability of direct transmission as a baseline for the comparison purpose. As is known, an intercept event occurs when the secrecy capacity becomes negative. Thus, the intercept probability of direct transmission is obtained from Eq. (5) as
\begin{equation}\label{equa13}
\begin{split}
P_{{\textrm{intercept}}}^{\textrm{direct}} = \Pr \left( {{C^{\textrm{direct}}_{sd}} < {C^{\textrm{direct}}_{se}}} \right) = \Pr \left( {|{h_{sd}}{|^2} < |{h_{se}}{|^2}} \right),
\end{split}
\end{equation}
where the second equation is obtained by using Eqs. (3) and (4). Considering that $|h_{sd}|^2$ and $|h_{se}|^2$ are independent exponentially distributed, we obtain a closed-form intercept probability expression of direct transmission as
\begin{equation}\label{equa14}
P_{{\textrm{intercept}}}^{\textrm{direct}} = \dfrac{{\sigma _{se}^2}}{{\sigma _{se}^2 + \sigma _{sd}^2}}=\dfrac{1}{{1 + \lambda _{de}^{ - 1}}} \cdot \dfrac{1}{{{\lambda _{de}}}},
\end{equation}
where $\sigma _{se}^2 = E(|{h_{se}}{|^2})$, $\sigma _{sd}^2 = E(|{h_{sd}}{|^2})$, and $\lambda_{de}=\sigma^2_{sd}/\sigma^2_{se}$ is the ratio of average channel gain from source to destination to that from source to eavesdropper, which is referred to as the main-to-eavesdropper ratio (MER) throughout this paper. It is observed from Eq. (14) that the intercept probability of direct transmission is independent of the transmit power $P$, which implies that the wireless security performance cannot be improved by increasing the transmit power. This motivates us to exploit cooperative relays to decrease the intercept probability and improve the physical-layer security.

\subsection{Traditional Max-Min Scheme}
This subsection presents the intercept probability analysis of traditional max-min scheme in Rayleigh fading channels. From Eq. (11), we obtain an intercept probability of the traditional max-min scheme as
\begin{equation}\label{equa15}
P_{{\textrm{intercept}}}^{\textrm{traditional}} = \Pr \left( {\mathop {\max }\limits_{i \in {\cal{R}}} C_{sid}^{\textrm{DF}} < C_{be}^{\textrm{DF}}} \right),
\end{equation}
where $C_{be}^{\textrm{DF}}$ denotes the channel capacity from the best relay to eavesdropper with DF relaying protocol. Similarly, assuming that ${h_{si}}$ and ${h_{id}}{\textrm{ }}(i=1, \cdots ,M)$ are identically and independently distributed and using the law of total probability, the intercept probability of traditional max-min scheme is given by
\begin{equation}\label{equa16}
P_{{\textrm{intercept}}}^{\textrm{traditional}} = \sum\limits_{m = 1}^M {\frac{1}{M}\Pr \left( {\mathop {\max }\limits_{i \in {\cal{R}}} \min \left( {|{h_{si}}{|^2},|{h_{id}}{|^2}} \right) < |{h_{me}}{|^2}} \right)} .
\end{equation}
Notice that $|{h_{si}}{|^2}$, $|{h_{id}}{|^2}$ and $|{h_{me}}{|^2}$ follow exponential distributions with means $\sigma^2_{si}$, $\sigma^2_{id}$ and $\sigma^2_{me}$, respectively. Letting $x=|{h_{me}}{|^2}$, we obtain Eq. (17) at the top of the following page,
\begin{figure*}
\begin{equation}\label{equa17}
\begin{split}
P_{{\textrm{intercept}}}^{\textrm{traditional}}&= \sum\limits_{m = 1}^M {\frac{1}{M}\int_0^\infty  {\prod\limits_{i = 1}^M {[1 - \exp ( - \frac{x}{{\sigma _{si}^2}} - \frac{x}{{\sigma _{id}^2}})]} \frac{1}{{\sigma _{me}^2}}\exp ( - \frac{x}{{\sigma _{me}^2}})dx} } \\
&= \sum\limits_{m = 1}^M {\frac{1}{M}\int_0^\infty  {\left( {1 + \sum\limits_{k = 1}^{{2^M} - 1} {{{( - 1)}^{|{{\mathcal{A}}_k}|}}\exp [ - \sum\limits_{i \in {{\mathcal{A}}_k}} {(\frac{x}{{\sigma _{si}^2}} + \frac{x}{{\sigma _{id}^2}})} ]} } \right)\frac{1}{{\sigma _{me}^2}}\exp ( - \frac{x}{{\sigma _{me}^2}})dx} } \\
&= \sum\limits_{m = 1}^M {\frac{1}{M}\left( {1 + \sum\limits_{k = 1}^{{2^M} - 1} {{{( - 1)}^{|{{\mathcal{A}}_k}|}}{[1 + \sum\limits_{i \in {{\mathcal{A}}_k}} {(\frac{{\sigma _{me}^2}}{{\sigma _{si}^2}} + \frac{{\sigma _{me}^2}}{{\sigma _{id}^2}})} ]^{- 1}}} } \right)}
\end{split}
\end{equation}
\end{figure*}
where the second equation is obtained by using the binomial expansion, ${\mathcal{A}}_k$ represents the $k$-th non-empty sub-collection of $M$ relays, and $|{\mathcal{A}}_k|$ represents the number of elements in set ${\mathcal{A}}_k$.

\subsection{Proposed Best Relay Selection Scheme}
This subsection derives a closed-form intercept probability expression of the proposed best relay selection scheme. According to the definition of intercept event, an intercept probability of proposed scheme is obtained from Eq. (12) as
\begin{equation}\label{equa18}
\begin{split}
P_{{\textrm{intercept}}}^{\textrm{proposed}} &= \Pr \left( {\mathop {\max }\limits_{i \in {\mathcal{R}}} C_i^{\textrm{DF}} < 0} \right) \\
&=\prod\limits_{i = 1}^M {\Pr \left\{ {\min (|{h_{si}}{|^2},|{h_{id}}{|^2}) < |{h_{ie}}{|^2}} \right\}}  ,
\end{split}
\end{equation}
where the second equation is obtained by using Eq. (10). Notice that random variables $|{h_{si}}{|^2}$, $|{h_{id}}{|^2}$ and $|{h_{ie}}{|^2}$ follow exponential distributions with means $\sigma^2_{si}$, $\sigma^2_{id}$ and $\sigma^2_{ie}$, respectively. Denoting $X=\min (|{h_{si}}{|^2}, |{h_{id}}{|^2})$, we can easily obtain the cumulative density function (CDF) of $X$ as
\begin{equation}\label{equa19}
{P_X}(X < x) = 1 - \exp ( - \frac{x}{{\sigma _{si}^2}} - \frac{x}{{\sigma _{id}^2}}),
\end{equation}
wherein $x\ge0$. Using Eq. (19), we have
\begin{equation}\label{equa20}
P_{{\textrm{intercept}}}^{\textrm{proposed}}  = \prod\limits_{i = 1}^M \dfrac{{\sigma _{id}^2\sigma _{ie}^2 + \sigma _{si}^2\sigma _{ie}^2}}{{\sigma _{id}^2\sigma _{ie}^2 + \sigma _{si}^2\sigma _{ie}^2 + \sigma _{si}^2\sigma _{id}^2}},
\end{equation}
which completes the closed-form intercept probability analysis of OAS scheme in Rayleigh fading channels.

\section{Numerical Results}

\begin{figure}
  \centering
  {\includegraphics[scale=0.55]{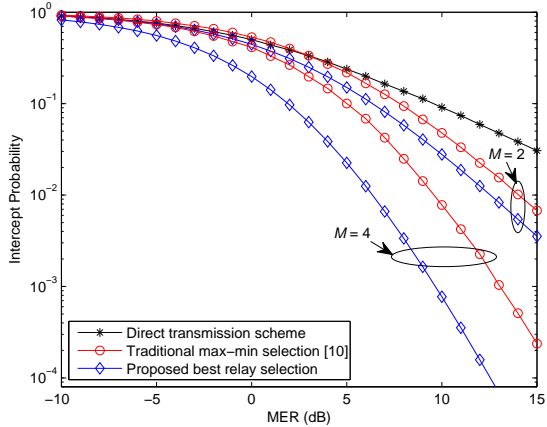}\\
  \caption{Intercept probability versus MER of various transmission schemes with $\alpha_{si}=\alpha_{id}=\alpha_{ie}=1$ wherein ${\alpha _{si}} = \sigma _{si}^2/\sigma _{sd}^2$, ${\alpha _{id}} = \sigma _{id}^2/\sigma _{sd}^2$ and ${\alpha _{ie}} = \sigma _{ie}^2/\sigma _{se}^2$.}\label{Fig2}}
\end{figure}

Fig. 2 shows the numerical intercept probability results of direct transmission, traditional max-min relay selection, and proposed best-relay selection schemes by plotting Eqs. (14), (17) and (20) as a function of MER. One can see from Fig. 2 that for both cases of $M=2$ and $M=4$, the intercept probability of proposed best relay selection scheme is always smaller than that of traditional max-min relay selection, showing the advantage of proposed best relay selection over traditional max-min scheme. Fig. 3 depicts the intercept probability versus the number of relays $M$ of the traditional max-min and proposed best relay selection schemes. It is observed from Fig. 3 that the proposed best relay selection scheme strictly performs better than the traditional max-min scheme in terms of the intercept probability. Fig. 3 also shows that as the number of relays $M$ increases, the intercept probabilities of both the traditional and proposed relay selection schemes significantly decrease. This means that increasing the number of relays can greatly improve the physical-layer security against eavesdropping attack.

\begin{figure}
  \centering
  {\includegraphics[scale=0.55]{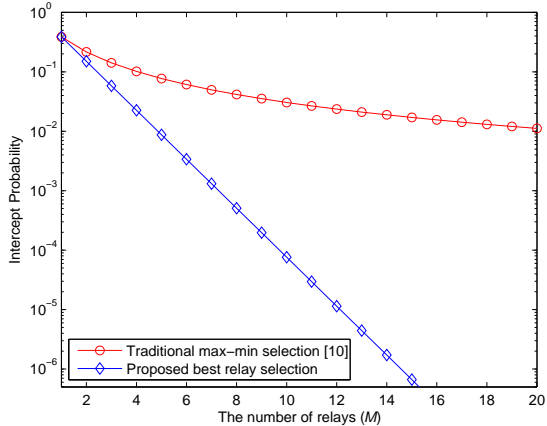}\\
  \caption{Intercept probability versus the number of relays $M$ of the traditional max-min relay selection and proposed best relay selection schemes with $\lambda_{de}=5{\textrm{dB}}$ and $\alpha_{si}=\alpha_{id}=\alpha_{ie}=1$ wherein ${\alpha _{si}} = \sigma _{si}^2/\sigma _{sd}^2$, ${\alpha _{id}} = \sigma _{id}^2/\sigma _{sd}^2$ and ${\alpha _{ie}} = \sigma _{ie}^2/\sigma _{se}^2$.}\label{Fig3}}
\end{figure}

\section{Conclusion}
In this paper, we investigated the physical-layer security in cooperative wireless networks with multiple DF relays and proposed the best relay selection scheme to improve wireless security against eavesdropping attack. We also examined the direct transmission and traditional max-min relay selection as benchmark schemes. We derived closed-form intercept probability expressions of the direct transmission, traditional max-min relay selection, and proposed best relay selection schemes. Numerical intercept probability results showed that the proposed best relay selection scheme always performs better than the direct transmission and max-min relay selection schemes. Moreover, as the number of relays increases, the max-min relay selection and proposed best relay selection schemes both significantly improve, which shows the advantage of exploiting multiple relays for the physical-layer security improvement.

\section{Acknowledgment}
The work presented in this paper is partially supported by the Auto21 Network of Centre of Excellence, Canada.

\clearpage

\end{document}